\documentclass[%
 reprint,
showpacs,preprintnumbers,
 amsmath,amssymb,
 aps,
]{revtex4-1}

\usepackage{graphicx}
\usepackage{dcolumn}
\usepackage{bm}
\usepackage{color}

\usepackage{amsmath}

\usepackage{hyperref}
\usepackage{soul}
\usepackage{ulem}

\begin{document}


\title{ Valley- and spin-switch effects in molybdenum disulfide superconducting spin valve}
\author{Leyla Majidi}
\email{leyla.majidi@ipm.ir}
\author{Reza Asgari}
\affiliation{School of Physics, Institute for Research in Fundamental Sciences (IPM), Tehran 19395-5531, Iran}

\date{\today}

\begin{abstract}
We propose a hole-doped molybdenum disulfide (MoS$_2$) superconducting spin valve (F/S/F) hybrid structure in which the Andreev reflection process is suppressed for all incoming waves with a determined range of the chemical potential in ferromagnetic (F) region and the cross-conductance in the right F region depends crucially on the configuration of magnetizations in the two F regions. Using the scattering formalism, we find that the transport is mediated purely by elastic electron cotunneling (CT) process in a parallel configuration and changes to the pure crossed Andreev reflection (CAR) process in the low-energy regime, without fixing of a unique parameter, by reversing the direction of magnetization in the right F region. This suggests both valley- and spin-switch effects between the perfect elastic CT and perfect CAR processes and makes the nonlocal charge current to be fully valley- and spin-polarized inside the right F region where the type of the polarizations can be changed by reversing the magnetization direction in the right F region. We further demonstrate that the presence of the strong spin-orbit interaction $\lambda$ and an additional topological term ($\beta$) in the Hamiltonian of MoS$_2$ result in an enhancement of the charge conductance of the CT and CAR processes and make them to be present for long lengths of the superconducting region. Besides, we find that the thermal conductance of the structure with a small length of the highly doped superconducting region exhibits linear dependence on the temperature at low temperatures whereas it enhances exponentially at higher temperatures. In particular, we demonstrate that the thermal conductance versus the strength of the exchange field ($h$) in F region displays a maximum value at $h<\lambda$, which moves towards larger exchange fields by increasing the temperature.
\end{abstract}
\maketitle
\section{\label{sec:intro}Introduction}
Many applications of quantum information require entanglement of quantum states~\cite{Einstei35,Bennett2000,Nielsen04}. In condensed matter physics, the controlled creating and detecting of entangled electrons is an ongoing challenge. S-wave superconductors have been proposed as natural sources for entangled electrons~\cite{Zeilinger99,Recher01}, as Cooper pairs consist of two electrons which are both spin- and momentum-entangled. The Cooper pair can be spatially deformed by means of an inverse crossed (nonlocal) Andreev reflection (CAR) process in superconducting heterostructures. In CAR process, an electron excitation and a hole excitation at two separate metallic leads are coupled by means of Andreev scattering processes at two spatially distinct interfaces. Therefore, it is worthwhile studying the properties of the CAR and controlling its magnitude.

Much theoretical~\cite{Byers95,Falci01,Chtchelkatchev03,Bignon04} and experimental~\cite{Beckmann04,Russo05,Cadden06} efforts have been devoted to the CAR process by which a superconducting (S) region emits two electrons into two normal metallic (N) leads such that they propagate in opposite directions and can be probed separately. Therefore, a straightforward way of observing CAR is nonlocal conductance measurements~\cite{Beckmann04,Russo05}. Unfortunately, in ordinary nonrelativistic systems, the small value of the CAR conductance is completely canceled by the conductance of another nonlocal process known as elastic electron cotunneling (CT) which does not involve Copper pairs and is therefore a parasitic process~\cite{Falci01}. Recently, it has been demonstrated that the atomically thin two-dimensional (2D) crystals such as graphene and silicene are possible areas for CAR processes~\cite{Cayssol08,Linder09,majidi12,majidi13,JWang12,Linder14} where the magnitude of the CAR conductance can be enhanced in normal/superconducting/normal (N/S/N) and ferromagnetic/superconducting/ferromagnetic (F/S/F) hybrid structures. Cayssol~\cite{Cayssol08} has predicted a pure CAR process without any valley- or spin-polarization in graphene-based N/S/N structure at the bias voltage $eV=\mu$ ($\mu$ is the chemical potential). Linder~{\it et al.}~\cite{Linder09} proposed a graphene-based superconducting spin valve (F/S/F) structure which creates a spin-polarized nonlocal current via the pure CAR (CT) process in the antiparallel (parallel) configuration of the magnetizations of F regions at a fixed chemical potential $\mu=h$ ($h$ is the exchange energy of the F region). A pure CAR process has also been shown in a magnetized zigzag graphene nanoribbon (MZR)/S/ZR junction with an even zigzag chain number for the ribbon~\cite{JWang12}. Furthermore, Linder \textit{et al.}~\cite{Linder14} have recently demonstrated a fully spin$\times$valley polarized pure CAR process (with $s\tau=-1$, where $s$ and $\tau$ indicate the spin and valley degrees of freedom) in silicene-based N/S/N structure which means that the nonlocal current is fully spin-polarized in each valley. Moreover, there has been many studies on the transport of charge, spin, and heat through the hybrid structures of normal, ferromagnetic and superconducting graphene~\cite{Hsu10,Salehi10,Wang11}.

Monolayer molybdenum disulfide (MoS$_2$) belongs to the family of 2D-layered transition metal dichalcogenides that have recently received significant attentions~\cite{Mattheis73}. Monolayer MoS$_2$ is a direct gap semiconductor with a band gap of $1.9$ eV which distinguishes it from its bulk and bilayer counterparts that are both indirect band gap materials with smaller value~\cite{Mak10,Splendiani10,Korn11}. As in graphene, the conduction and valence band edges consist of two degenerate valleys $ (K, K') $ located at the corners of the hexagonal Brillouin zone. What sets MoS$_2$ much more interesting is the presence of a strong spin-orbit coupling (originating from the heavy metal atoms) that produces a large spin splitting of the valence band~\cite{Xiao12,Zeng12,Cao12} and makes the monolayer MoS$_2$ a convenient platform to explore spin physics and spintronics applications. In addition, MoS$_2$ offers the possibility for a coupling of the spin and valley physics, which is owing to the strong spin-orbit coupling and the broken inversion symmetry. Since the two inequivalent valleys in the monolayer MoS$_2$ are separated in the Brillouin zone by a large momentum, in the case of the absence of short-range interactions, intervalley scattering~\cite{lu} should be negligible and thus the valley index becomes a new quantum number. Therefore, manipulating the valley quantum number can produce new physical effects. Recently, many measurements have been performed to characterize the optical and transport properties of the monolayer MoS$_2$~\cite{Splendiani10,Radisavljevic11,Wang12,Sallen12}. For practical applications in electronic devices, the single-layer and multi-layer MoS$_2$ can be $n$- or $p$-type doped to generate desirable charge carriers~\cite{Radisavljevic11, Fontana13}. At high carrier concentrations and in the presence of the high-$\kappa$ dielectrics, MoS$_2$ has been shown to undergo a superconducting transition, with a doping-dependent critical temperature~\cite{Gupta91, Takagi12, Ye12, Roldan13}. A ferromagnetic behavior has also been reported in MoS$_2$, and it has been related to edges~\cite{Zhang07,Li09,Mathew12,Ma12,Tongay}, doping with transition metals~\cite{Mishra13}, or to the existence of defects~\cite{Vojvodic09,Ataca11}.

In this paper, we propose a monolayer molybdenum disulfide superconducting spin valve (F/S/F) structure in which a pure CAR signal can be generated without any contamination from the elastic CT process. Importantly, in contrast to the aforementioned proposals in graphene and silicene, there is no fixing of a unique parameter to have a pure CAR process in the proposed MoS$_2$-based structure and it creates a fully valley- and spin-polarized nonlocal current for a wide range of the chemical potential of the F region. Within the scattering formalism, we find that for all incoming electrons to the $p$-doped molybdenum disulfide F/S/F structure with a determined range of the chemical potential $\mu_{F}$, the Andreev reflection (AR) process, conversion of the electron into the hole excitation at a N/S interface~\cite{Andreev64}, is suppressed and the transport is mediated purely by the CT process in parallel alignment of magnetizations and the CAR process in antiparallel configuration. This suggests valley- and spin-switching effects between the perfect CT and perfect CAR processes and makes the nonlocal charge current to be valley- and spin-polarized inside the right F region. Furthermore, we demonstrate that the presence of the strong spin-orbit interaction and a topological term ($\beta$) in the Hamiltonian~\cite{Rostami13} of MoS$_2$ enhance the charge conductance of the CT and CAR processes, respectively, in the parallel and antiparallel configurations and make them to decay slowly with the length of the S region. Our results reveal the potential of the proposed structure for application in valley-based electronics.

For the sake of completeness, we investigate the thermal transport characteristics of the proposed structure. We find that for MoS$_2$-based F/S/F structures with small lengths of the highly doped superconducting region, the thermal conductance displays linear dependence on the temperature at low temperatures and exponential form at high temperatures. The thermal conductance can also be increased or decreased with respect to the temperature, depending on the strength of the exchange field in F region. We further demonstrate that in contrast to the graphene-based F/S/F structure~\cite{Salehi10}, the thermal conductance displays a maximum value with respect to the exchange field at $h<\lambda$, which moves towards larger exchange fields by increasing the temperature.

This paper is organized as follows. In Sec.~\ref{sec:level1}, we present the proposed structure and establish the theoretical framework which will be used
to calculate the local and nonlocal conductances. The numerical results for the charge conductance of the CT and CAR processes
and the thermal conductance in the proposed MoS$_2$-based F/S/F hybrid structure are presented in Sec.~\ref{sec:level2}. Finally, a brief summary of results is given in Sec.~\ref{sec:level3}.
\begin{figure}[t]
\begin{center}
\includegraphics[width=3.4in]{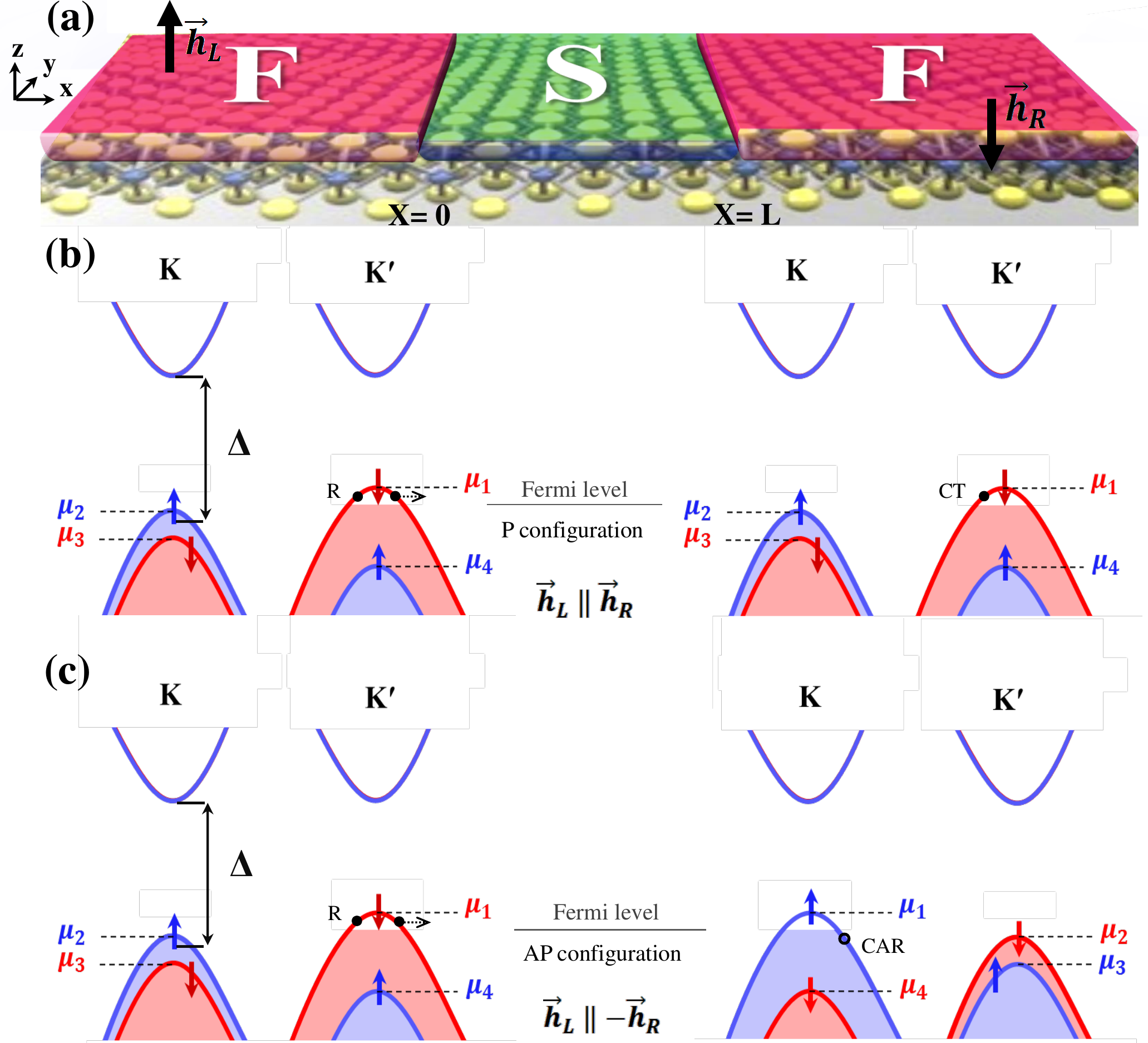}
\caption{\label{Fig:1} (Color online) (a) Schematic illustration of the proposed superconducting spin valve in a $p$-doped monolayer molybdenum disulfide with antiparallel configuration: The left and right regions are proximity induced ferromagnetic (F) regions, respectively, with the exchange fields $\bm{h}_L=h \hat{z}$ and $\bm{h}_R=-\bm{h}_L$, and the intermediate region is in the superconducting (S) state (caused by proximity to a top S electrode). (b)-(c) The dispersion relation in momentum space of $p$-doped F regions with the magnitude of the exchange field $h<\lambda$ for parallel (P) and antiparallel (AP) alignments of magnetizations at the $K$ and $K'$ valleys. The conduction and the valence bands are separated by a large band gap $\Delta=1.9$ eV. The energies of the valence band edges for different spin-subbands of two valleys are $\mu_1=-\Delta/2+\lambda+h$, $\mu_2=-\Delta/2+\lambda-h$, $\mu_3=-\Delta/2-\lambda+h$, and $\mu_4=-\Delta/2-\lambda-h$, which are measured from the center of the gap $\Delta$ (zero-energy point).}
\end{center}
\end{figure}

\section{\label{sec:level1}Model and Theory}

We consider a wide $p$-doped monolayer molybdenum disulfide F/S/F junction which constitutes a superconducting spin valve structure and is shown schematically in Fig.~\ref{Fig:1}(a). A s-wave superconducting top electrode covers the central region ($0<x<L$), creating a proximity induced superconducting correlations between the left ($x<0$) and right ($x>L$) ferromagnetic (F) regions. As a consequence, a superconducting gap $\Delta_S$ opens in the central region (the superconducting order parameter $\Delta_S$ is taken to be real and constant for s-wave pairing). The ferromagnetism is also assumed to be induced by means of the proximity effect to the F lead with desired properties. Such a F region in graphene can be produced by using an insulating ferromagnetic substrate, or by adding F metals or magnetic impurities on top of the graphene sheet~\cite{Swartz12,Dugaev06,Tombros07,Yazyev10}. The induced exchange field in the left F region is $\bm{h}_L=h \hat{z}$, while it is $\bm{h}_R=\pm\bm{h}_L$ in the right F region for the parallel (antiparallel) configuration. In experiments, the two parallel (P) and antiparallel (AP) configurations of the proximity magnetizations can be adjusted by externally applied magnetic fields~\cite{Meier00,Meier01}. The role of the finite-size effect for a nanoribbon MoS$_2$ will not be addressed in the present work. We note that the finite-size effects may lead to different results with respect to the bulk molybdenum disulfide case as it has been argued regarding graphene structure~\cite{Rainis09,Haugen10}.

We use Dirac-Bogoliubov-de Gennes (DBdG) equation~\cite{beenakker06,Majidi14} which describes the superconducting correlations between relativistic electrons and holes with opposite spins and different valley indices (for more information on the derivation of the DBdG equation for MoS$_2$, see Appendix). In the presence of an exchange interaction, the DBdG equation has the form
\par
\begin{equation}
\label{DBdG}
\hspace{-0.5cm}\left(
\begin{array}{cc}
H_{\tau}-sh & \Delta_S \\
\Delta_{S}^{\ast}& -(H_{\tau}-{\bar{s}}h)
\\
\end{array}
\right)
\left(
\begin{array}{c}
u_{s,\tau}\\
v_{\bar{s},\bar{\tau}}
\end{array}
\right)
=\varepsilon\left(
\begin{array}{c}
u_{s,\tau}\\
v_{\bar{s},\bar{\tau}}
\end{array}
\right),\nonumber\\
\end{equation}
where
\begin{eqnarray}
\label{H}
H_{\tau}&=&v_{\rm F}(\bm{\sigma}_{\tau}.\bm{p})+\frac{\Delta}{2}\sigma_{z}+\lambda s \tau\ (\frac{1-\sigma_z}{2})\nonumber\\
&+&\frac{\bm{p}^2}{4m_0}(\alpha+\beta\sigma_z)+U(\bm{r})-\mu,
\end{eqnarray}
is the modified Dirac Hamiltonian~\cite{Rostami13} of a monolayer MoS$_2$ for spin $s=\pm 1$ ($\bar{s}=-s$) and valley $\tau=\pm 1$ ($\bar{\tau}=-\tau$) in which the energy gap $\Delta= 1.9$ eV, spin-orbit coupling constant $\lambda= 0.08$ eV, $v_{\rm F}=0.53\times10^6$ m/s is the Fermi velocity, $m_0$ is the bare electron mass, $\alpha = 0.43$, and $\beta=2.21$. The $\alpha$ term is originated from the difference between electron and hole masses recently reported by using \textit{ab initio} calculations~\cite{Peelaers12} and in addition, the $\beta$ term leads to a new topological characteristic~\cite{rostami_opt}. The exchange field $h$, and the electrostatic potential $U(\bm{r})$, are taken to be zero and $-U_0$, respectively, in the S region and $U(\bm{r})=0$ in both F regions. Here, $\varepsilon$ is the excitation energy and the electron and hole wave functions, $u_{s,\tau}$ and $v_{\bar{s},\bar{\tau}}$, are two-component spinors of the form $(\psi_c,\psi_v)$, where $c$ and $v$ denote the conduction and valence bands, respectively, and $\bm{\sigma}_{\tau}=(\tau\sigma_x,\sigma_y,\sigma_z)$
is the vector of the Pauli matrices acting on the two conduction and valence bands.

Let us assume that a spin-$s$ valley-$\tau$ electron is injected from the left F region with the excitation energy less than the superconducting gap, $\varepsilon< \Delta_S$. There are four possible scattering mechanisms. The electron can be reflected normally (R) or tunneled through the sample via the elastic CT process. To enter the superconducting condensate, it needs a partner of the opposite spin and different valley index, which ejects a spin-$\bar{s}$ hole from valley-$\bar{\tau}$ in either the left or right F region via the local AR or CAR process, respectively (for more details on Andreev reflection in monolayer MoS$_2$, see Ref.~\onlinecite{Majidi14}). Denoting the amplitudes of the normal R, AR, CAR, and CT processes $r_e^{s,\tau}$, $r_h^{s,\tau}$, $t_h^{s,\tau}$ and $t_e^{s,\tau}$, respectively, the total wave functions inside the left and right F regions can be written as
\par
\begin{widetext}
\begin{eqnarray}
\label{leftF}
\psi_{L}&=&\psi_s^{e+}+r_e^{s,\tau}\ \psi_s^{e-}+r_{h}^{s,\tau}\ \psi_{\bar{s}}^{h-}\nonumber\\\nonumber\\
&=&\frac{1}{\sqrt{u_e}}\ e^{-ik_{e}^{s,\tau}\tau x} e^{iqy}
\left(
\begin{array}{c}
e^{i\tau\theta_e/2}\\
-a_e\ \tau\ e^{-i\tau\theta_e/2}\\
0\\
0
\end{array}
\right)+\frac{r_e^{s,\tau}}{\sqrt{u_e}}\ e^{ik_e^{s,\tau}\tau x} e^{iqy}
\left(
\begin{array}{c}
e^{-i\tau\theta_e/2}\\
a_e\ \tau\ e^{i\tau\theta_e/2}\\
0\\
0
\end{array}
\right)+\frac{r_{h}^{s,\tau}}{\sqrt{u_h}}\ e^{-ik_h^{s,\tau}\tau x} e^{iqy}
\left(
\begin{array}{c}
0\\
0\\
e^{i\tau\theta_h/2}\\
-a_h\ \tau\ e^{-i\tau\theta_h/2}
\end{array}
\right),\nonumber\\\\
\psi_{R}&=&t_e^{s,\tau}\ \psi_s^{'e+}+t_h^{s,\tau}\ \psi_{\bar{s}}^{'h+}\nonumber\\
&=&\frac{t_e^{s,\tau}}{\sqrt{u'_e}}\ e^{-i{k}_e^{'s,\tau}\tau x} e^{iqy}
\left(
\begin{array}{c}
e^{i\tau\theta'_e/2}\\
-a'_e\ \tau\ e^{-i\tau\theta'_e/2}\\
0\\
0
\end{array}
\right)+\frac{t_h^{s,\tau}}{\sqrt{u'_h}}\ e^{i{k}_h^{'s,\tau}\tau x} e^{iqy}
\left(
\begin{array}{c}
0\\
0\\
e^{-i\tau\theta'_h/2}\\
a'_h\ \tau\ e^{i\tau\theta'_h/2}
\end{array}
\right),\nonumber\\
\end{eqnarray}
where $\psi_s^{(')e\pm}$ and $\psi_{\bar{s}}^{(')h\pm}$ are the solutions of the DBdG equation for valence band electrons and holes of the left (right) $p$-doped F region with the chemical potential $\mu_{F} = \mu$ ($\mu<0$ is measured from the center of the gap, $\Delta$) at a given energy $\varepsilon$ and transverse wave vector $q$ with the energy-momentum relation that can be obtained (from the DBdG equation with $\Delta_S =0$) by solving the following equation,
\begin{equation}\label{general}
\left(\frac{\hbar^2|\bm{k}_{e(h)}^{('){s,\tau}}|^2}{4 m_0} (\alpha +\beta)+\frac{\Delta }{2}-\mu_{F}-s(\bar{s})h_R\mp\epsilon \right)\ \left(\frac{\hbar^2 |\bm{k}_{e(h)}^{('){s,\tau}}|^2 }{4 m_0}(\alpha -\beta )+\lambda s  \tau -\frac{\Delta }{2}-\mu_{F}-s(\bar{s})h_R\mp\epsilon \right)-\hbar^2 v_{\rm F}^2 |\bm{k}_{e(h)}^{('){s,\tau}}|^2=0.
\end{equation}
$u_{e(h)}^{(')}=\hbar |\bm{k}_{e(h)}^{('){s,\tau}}|\cos({\tau\theta_{e(h)}^{(')}})\ [\alpha+\beta+{a_{e(h)}^{(')}}^2\ (\alpha-\beta)]/4m_0 v_{\rm F}+a_{e(h)}^{(')}\cos({\tau\theta_{e(h)}^{(')}})$,
$a_{e(h)}^{(')}=\hbar v_{\rm F} |\bm{k}_{e(h)}^{('){s,\tau}}|/[\mu_{F}+s(\bar{s})h_R\pm\varepsilon+\Delta/2-\lambda s \tau-\hbar^2|\bm{k}_{e(h)}^{('){s,\tau}}|^2(\alpha-\beta)/4 m_0]$, $h_R=\pm h$ is the exchange field in the right F region for the P(AP) configuration and $\theta_{e(h)}^{(')}=\arcsin({q/|\bm{k}_{e(h)}^{('){s,\tau}}|})$ indicates the angle of the propagation of the electron (hole). Also, the two propagation directions along the $x$ axis are denoted by $\pm$ in $\psi_{s}^{(')e\pm}$ and $\psi_{\bar{s}}^{(')h\pm}$.
\par
Furthermore, the total wave function inside the S region can be written as,
\par
\begin{eqnarray}
\psi_{S}&=&t_1\ \psi^{S+}+t_2\ \psi^{S'+}+t_3\ \psi^{S-}+t_4\ \psi^{S'-}\nonumber\\\nonumber\\
&=&\Sigma_{j=1}^4\ t_j\ e^{ik_j\tau x} e^{iqy}
\left(
\begin{array}{c}
b_j\\
-a_j\ c_j\\
1\\
-a_j
\end{array}
\right),
\end{eqnarray}
where $\psi^{S\pm}$ and $\psi^{S'\pm}$ are the solutions of the DBdG equation for right (left) going Dirac-Bogoliubov quasiparticles (mixed electron-hole excitations) inside the S region with
\begin{eqnarray}
a_j&=&\frac{m_j+{\Delta_S^2}}{\hbar v_{\rm F} (\tau k_j-iq)\ m'_j},\nonumber\\
b_j&=&\frac{m'_j\ \Delta_S}{-m'_j\ (\frac{\Delta}{2}+\frac{\hbar^2 k_{Sj}^2 (\alpha+\beta)}{4m_0}-\mu_S-\varepsilon)+ m_j+{\Delta_S^2}},\nonumber\\
c_{j}&=&\frac{\Delta_S\ (m_{j}+{\Delta_S^2})}{\hbar^2v_{\rm F}^2{k}_{Sj}^2 m'_{j}-(m_{j}+{\Delta_S^2})(-\frac{\Delta}{2}+\lambda s \tau+\frac{\hbar^2 k_{Sj}^2 (\alpha-\beta)}{4m_0}-\mu_S-\varepsilon)},\nonumber
\end{eqnarray}
$k_{1(3)}=\mp k_0+i k'\tau$, $k_{2(4)}=-k_{1(3)}$, $k_{Sj}=\sqrt{{k_j}^2+q^2}$, $\mu_S=\mu+U_0$, $m_j=[{\Delta}/{2}+\hbar^2 k_{Sj}^2 (\alpha+\beta)/4m_0-\mu_S]^2-\varepsilon^2+\hbar^2v_{\rm F}^2{k}_{Sj}^2$ and $m'_j=\hbar^2 k_{Sj}^2 \alpha/2m_0-2\mu_S+\lambda s \tau$. The momentum $k_{Sj}$ of qausiparticles in the S region are the solutions of the energy-momentum relation, which can be obtained (from the DBdG equation) by solving the equation
\begin{eqnarray}
\label{ks}
&&\hspace{-5mm}\varepsilon^4 - B\ \varepsilon^2 + C = 0,\nonumber\\
&&\hspace{-5mm}B=\left(\frac{\hbar^2 k_{Sj}^2 }{4 m_0}(\beta -\alpha )-\lambda s \tau +\frac{\Delta }{2}+\mu_S\right)^2+\left(\frac{\hbar^2 k_{Sj}^2 }{4m_0}(\alpha +\beta )+ \frac{\Delta }{2} - \mu_S\right)^2+2\ (\hbar^2 v_{\rm F}^2 k_{Sj}^2+ \Delta_S^2),\nonumber\\
&&\hspace{-5mm}C=\left([\frac{\hbar ^2 k_{Sj}^2 }{4 m_0}(\alpha +\beta )+\frac{\Delta }{2}-\mu_S]^2+\hbar ^2 v_{\rm F}^2 k_{Sj}^2+\Delta_S^2\right) \left([\ \frac{\hbar ^2 k_{Sj}^2}{4 m_0} (\beta-\alpha)- \lambda s \tau +\frac{\Delta }{2}+\mu_S ]^2+\hbar ^2 v_{\rm F}^2 k_{Sj}^2+\Delta_S^2\right)\nonumber\\
&&\hspace{0.2cm}-\hbar ^2 v_{\rm F}^2 k_{Sj}^2 \left(\frac{\hbar ^2 k_{Sj}^2}{2m_0}\ \alpha + \lambda s  \tau -2 \mu_S\right)^2.
\end{eqnarray}
\par
\end{widetext}
We calculate, matching the wave functions at the two interfaces ($x=0,\ x=L$), the normal and Andreev reflection amplitudes in the left F region and the transmission amplitudes of the electron and the hole into the right F region of the both P and AP configurations. Having calculated the reflection and transmission amplitudes and replacing them in the BTK formula~\cite{Blonder82}, we can calculate the charge conductance of the AR, CT and CAR processes at zero temperature as
\begin{eqnarray}
\label{G_AR}
&&\hspace{-4mm}G_{AR}=\sum_{s,\tau=\pm1}G_{0}^{s,\tau}\int_{0}^{\theta_{AR}}(1-|r_e^{s,\tau}|^2+|r_{h}^{s,\tau}|^2)\cos{\theta_e}\ d\theta_e,\nonumber\\\\
\label{G_CT}
&&\hspace{-4mm}G_{CT}=\sum_{s,\tau=\pm1}G_{0}^{s,\tau}\int_{0}^{\theta_{CT}}|t_{e}^{s,\tau}|^2\cos{\theta_e}\ d\theta_e,\\
\label{G_CAR}
&&\hspace{-4mm}G_{CAR}=\sum_{s,\tau=\pm1}G_{0}^{s,\tau}\int_{0}^{\theta_{CAR}}|t_{h}^{s,\tau}|^2\cos{\theta_e}\ d\theta_e,
\end{eqnarray}
where we introduce $G_{0}^{s,\tau}={e^2} N_{s,\tau}(eV)/{2\pi\hbar}$ as the spin-$s$ valley-$\tau$ normal state charge conductance and $\ N_{s,\tau}(\varepsilon)={W|\bm{k}_e^{s,\tau}|}/{\pi}$ as the number of transverse modes in a sheet of monolayer MoS$_2$ of width $W$. Here, $\theta_{AR}=\arcsin({|\bm{k}_h^{s,\tau}|}/{|\bm{k}_e^{s,\tau}|})$, $\theta_{CT}=\arcsin({|\bm{k}_e^{'s,\tau}|}/{|\bm{k}_e^{s,\tau}|})$, and $\theta_{CAR}=\arcsin({|\bm{k}_h^{'s,\tau}|}/{|\bm{k}_e^{s,\tau}|})$ are, respectively, the critical angle of incidence for the AR, CT, and CAR processes above which the corresponding waves do not contribute to any transport of charge. Also, we put $\varepsilon=eV$ at zero temperature.
\par
In order to complete the investigation of the transport properties of the proposed F/S/F structure, we calculate the thermal conductance of the device by incorporating only the low-energy excitations and assuming a temperature gradient through the junction, as follows~\cite{Bardas95,Yokoyama08}
\begin{eqnarray}
\label{kappa}
\kappa&=&A'\sum_{s,\tau=\pm1}\int_0^{\infty} \int_{-\pi/2}^{\pi/2}d\varepsilon\ d\theta_e \cos{\theta_e}\  \frac{\varepsilon^2\ |\bm{k}_{e}^{s,\tau}(\varepsilon)|}{(k_B T)^2{\cosh^2(\frac{\varepsilon}{2k_B T})}}\nonumber\\
&&(1-|r_e^{s,\tau}|^2-Re(\frac{\cos{\theta_h}}{\cos{\theta_e}})\ |r_{h}^{s,\tau}|^2),
\end{eqnarray}
where $A'=k_B/8\pi^2\hbar$. We note that, in contrast to the valley degeneracy in graphene, the contribution of each valley to the charge and thermal conductances must be computed separately.
\par
\begin{figure}[t]
\begin{center}
\includegraphics[width=3.3in]{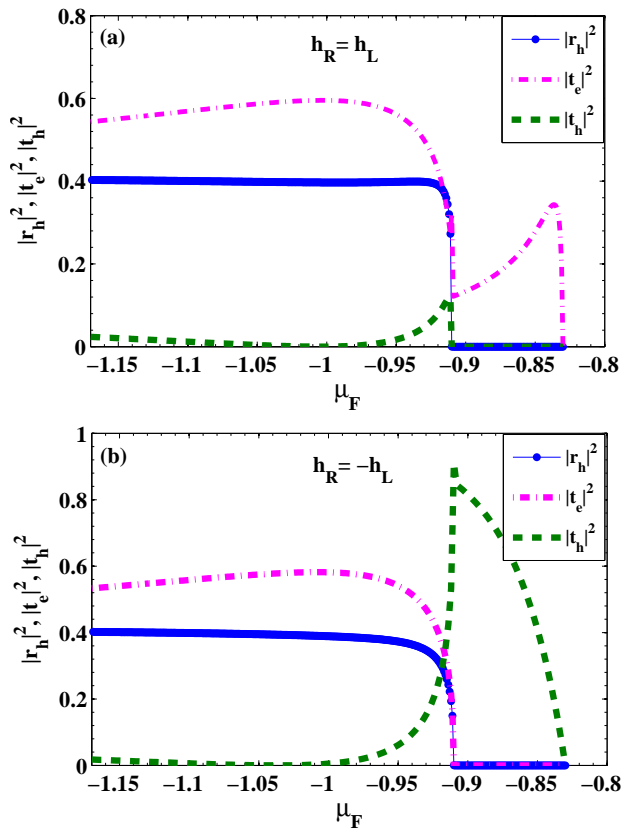}
\end{center}
\caption{\label{Fig:2} (Color online) Probabilities of the AR ($|r_h|^2$), CT ($|t_e|^2$) and CAR ($|t_h|^2$) processes as a function of the chemical potential of the F region, $\mu_F$, for normal incidence ($\theta_e=0$) to the MoS$_2$-based F/S/F structure with (a) parallel ($\bm{h}_R=\bm{h}_L$) and (b) antiparallel ($\bm{h}_R=-\bm{h}_L$) alignments of magnetizations, when $h_L=0.5 \lambda$, $L/\xi_S=0.5$, $\mu_S = -1$ eV, and $eV/\Delta_S = 0$.}
\end{figure}
\begin{figure}[]
\begin{center}
\includegraphics[width=3.4in]{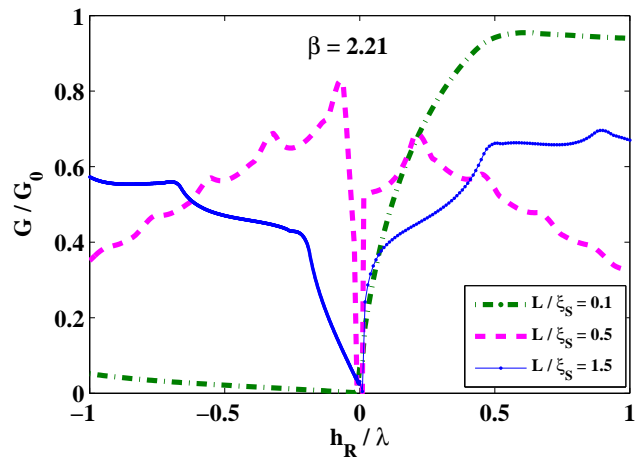}
\end{center}
\caption{\label{Fig:3} (Color online) Cross-conductance of the superconducting spin valve structure versus the exchange field of the right F region $\bm{h}_R/\lambda$ (in units of the spin-orbit coupling constant $\lambda$) for three different lengths of the S region, when $\alpha=0$, $\beta=2.21$, $eV/\Delta_S = 0$, $\mu_{F}=-\Delta/2+\lambda+eV-0.001$, $\mu_S = 1.1\ \mu_{F} = -1$ eV and $\Delta_S = 0.01$ eV.}
\end{figure}

\section{\label{sec:level2}NUMERICAL RESULTS and DISCUSSIONS}
\begin{figure}[t]
\begin{center}
\includegraphics[width=3.45in]{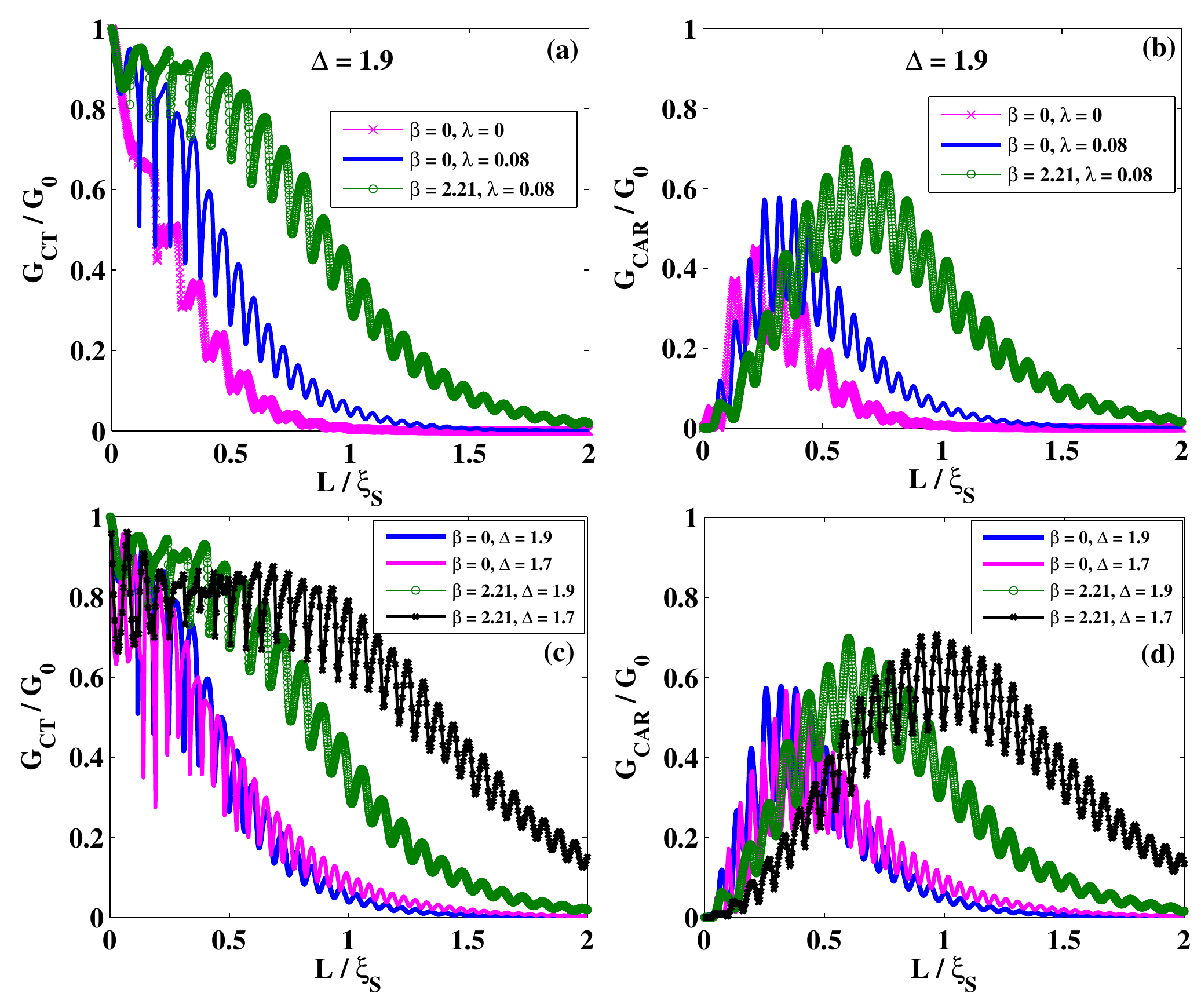}
\end{center}
\caption{\label{Fig:4} (Color online) The behavior of the charge conductance of the pure CT (left panel) and pure CAR (right panel) processes, respectively, in the parallel and antiparallel alignments of magnetizations versus the length of the S region $L/\xi_S$ (in units of the superconducting coherence length $\xi_S = \hbar v_{\rm F} / \Delta_S$) for gapped graphene ($\lambda=0, \beta=0$) and MoS$_2$ ($\lambda=0.08$) with $\alpha=0$ and $\beta=0$ and $2.21$, when $\Delta = 1.9$ eV (a)-(b) and for MoS$_2$ with $\Delta = 1.9$, and $1.7$ eV, $\alpha=0$ and $\beta= 0$ and $2.21$ (c)-(d), when $\Delta_S = 0.01$ eV, $\mu_{F}=-\Delta/2+\lambda-h+eV+0.001$, $\mu_S = 1.1\ \mu_{F} = -1 $ eV, $h = 0.5 \lambda$ and $eV / \Delta_S = 0$.}
\end{figure}
\begin{figure}[t]
\begin{center}
\includegraphics[width=3.4in]{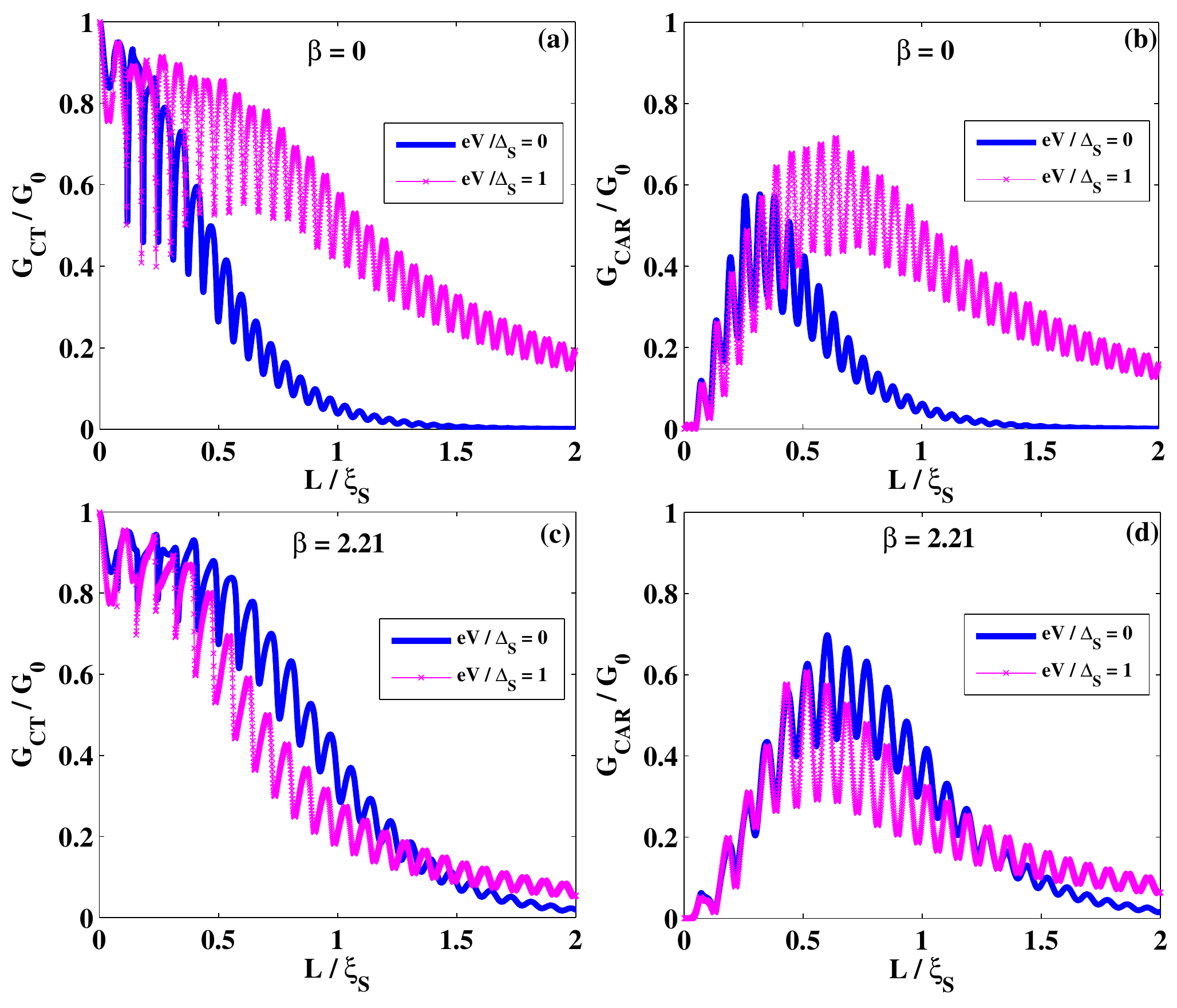}
\end{center}
\caption{\label{Fig:5} (Color online) Plots of the charge conductance of the CT (left panel) and CAR (right panel) processes, respectively, in the parallel and antiparallel configurations versus $L/\xi_S$ for two values of the bias voltage $eV / \Delta_S = 0$ and $1$, in MoS$_2$-based structure with $\alpha=0$ and $\beta=0$ (a)-(b) and $2.21$ (c)-(d), when $\Delta = 1.9$ eV, $\mu_{F}=-\Delta/2+\lambda-h+eV+0.001$, $\mu_S = 1.1\ \mu_{F} = -1 $ eV, $\Delta_S = 0.01$ eV and $h = 0.5 \lambda$.}
\end{figure}
In this section, we present our numerical results for the charge and thermal conductances, obtained by using the numerical scattering amplitudes $r_e^{s,\tau}$, $r_h^{s,\tau}$, $t_e^{s,\tau}$, $t_h^{s,\tau}$, and Eqs. (\ref{G_AR}), (\ref{G_CT}), (\ref{G_CAR}), and (\ref{kappa}). In the following, we scale the length of the superconducting region, $L$, in units of the superconducting coherence length $\xi_S=\hbar v_{\rm F}/\Delta_S$ ($\Delta_S$ is the zero-temperature superconducting order parameter) and all of the energies $\mu_S$, $\mu_F$, $\Delta$, $\lambda$ and $\Delta_S$ are in units of electron volt, eV.

\subsection{\label{subsecA}Charge Conductance}
At first, we evaluate the charge conductance of the proposed F/S/F structure at zero temperature. We consider the chemical potential $\mu_{F}$ in the range of $\mu_2 <\mu_{F}-eV\leq \mu_1$ ($\mu_1=-\Delta/2+\lambda+h$, $\mu_2=-\Delta/2+\lambda-h$) in which the contribution of the CT and CAR processes in the cross-conductance are separated directly by a simple valley- and spin-switch effects, and accordingly these processes are respectively present in the P and AP alignments of magnetizations of F regions. We have found that the presence of the mass asymmetry term ($\alpha$) in the Hamiltonian of MoS$_2$ has no significant effect on the charge conductance of the CT and CAR processes (similar behavior has been demonstrated for AR process in Ref.~[\onlinecite{Majidi14}]). Therefore, we set $\alpha=0$ in our investigations.

A schematic structure of the system is shown in Fig.~\ref{Fig:1} where the dispersion relation in momentum space of $p$-doped F regions of the proposed F/S/F structure with the magnitude of the exchange field $h<\lambda$ are illustrated for the parallel [Fig.~\ref{Fig:1}(b)] and antiparallel [Fig.~\ref{Fig:1}(c)] configurations at the two $K$ and $K'$ valleys. The chemical potentials $\mu_i\ (i=1-4)$ define the energies of the valence band edges for different spin subbands of two valleys. The incoming quasiparticles from the left F region with the chemical potential $\mu_2 <\mu_{F}-eV\leq \mu_1$ are completely dominated by the electrons with spin-down ($s=-1$) from $K'$ ($\tau=-1$) valley. Since there is no longer hole state with opposite spin ($s=1$) and different valley index ($\tau=1$) in the left F region, the AR process is suppressed and the incident electron can be normally reflected (R) as an electron into the left F region. In the P configuration, the incident electron can be tunneled through the sample via the elastic CT process while it can not be crossed Andreev reflected as a hole, since there is no hole state with $s=\tau=1$. This means that the transport is purely governed by the CT process in the P alignment of magnetizations. In the case of the AP configuration, the incident electron can be crossed Andreev reflected (CAR) as a hole with opposite spin ($s=1$) and different valley index ($\tau=1$) into the right F region, which means that the transport is mediated purely by the CAR process in the AP configuration.

These results can be also described from Fig.~\ref{Fig:2} where the probability of the AR, CT, and CAR processes in the P and AP configurations are plotted in terms of the chemical potential of the F region, $\mu_F$. For $\mu_2 <\mu_{F}-eV\leq \mu_1$ ($\mu_1=-0.83$ eV, $\mu_2=-0.91$ eV), the AR process is suppressed in both configurations and there is no probability to have the CAR and CT processes, respectively, in the P and AP configurations, while all of the scattering processes are possible for $\mu_{F}-eV\leq \mu_2$. Therefore, we obtain that the transport of the incoming electrons with a wide range of $\mu_F$ is mediated purely by the CT process in the P configuration and changes to the pure CAR process in the low-energy regime, by reversing the direction of the magnetization in the right F region. This suggests both valley- and spin-switch effects between the perfect elastic CT (transmission of electron with $s=-1$ and $\tau=-1$) and perfect CAR (transmission of hole with $s=1$ and $\tau=1$) processes for all applied subgap bias voltages by reversing the magnetization direction of the right F region [see Figs. \ref{Fig:1}(b)-\ref{Fig:1}(c)]. Moreover, the nonlocal charge current is fully valley- and spin-polarized inside the right F region of the MoS$_2$-based superconducting spin valve structure and the type of the polarizations can be changed by reversing the magnetization direction in the right F region. Furthermore, the behavior of the cross-conductance in the right F region is shown in Fig.~\ref{Fig:3} in terms of the exchange field of the right F region, $h_R$, for three different lengths of the S region, where the pure CAR process (for $h_R=-h_L<0$) is fully changed in the pure CT process (for $h_R=h_L>0$) by changing the configuration of the magnetization directions from AP to P. The length dependence of the CT and CAR conductances will be explained in details in the following paragraphs. We note that the existence of a fully valley- and spin-polarized nonlocal charge current via the pure CAR (CT) process, without fixing of any parameter and for a wide range of the chemical potential of the F region, is the advantage of the proposed MoS$_2$-based superconducting spin valve structure over the graphene- and silicene-based structures~\cite{Cayssol08,Linder09,majidi12,majidi13,JWang12,Linder14}. Therefore, we find that for all incoming electrons with $\mu_2 <\mu_{F}-eV\leq \mu_1$, the AR process is suppressed and the cross-conductance in the right F region depends crucially on the configuration of magnetizations in the two F regions.

In Fig.~\ref{Fig:4}, we plot the charge conductance of the CT (left panel) and CAR (right panel) processes, respectively in the parallel ($G_{CAR}\rightarrow 0$) and antiparallel ($G_{CT}\rightarrow 0$) alignments of magnetizations, as a function of the length of the S region $L/\xi_S$ for gapped graphene ($\lambda=0, \beta=0$) and MoS$_2$ ($\lambda=0.08$) with $\alpha=0$ and $\beta=0$ and $2.21$, when $\Delta = 1.9$ eV [Figs.~\ref{Fig:4}(a)-\ref{Fig:4}(b)] and for MoS$_2$ with $\Delta = 1.9$, and $1.7$ eV, $\alpha=0$ and $\beta= 0$ and $2.21$ [Figs. \ref{Fig:4}(c)-\ref{Fig:4}(d)], when $\Delta_S = 0.01$ eV, $\mu_{F}=-\Delta/2+\lambda-h+eV+0.001$, $\mu_S = 1.1\ \mu_{F} = -1 $ eV, $h = 0.5 \lambda$ and $eV / \Delta_S = 0$. We choose the chemical potential $\mu_{F}$ close to the energy of the valence band edge for $s=\tau=1$ spin-subband ($\mu_2$) because the conductance of the CT and CAR processes increase with $\mu_{F}$. The charge conductance of the CT and CAR processes are normalized to the normal state charge conductance $G_0=\sum_{s,\tau=\pm1}G_{0}^{s,\tau}$. It is seen that the CT process is favored for short junctions $L\ll \xi_{S}$, while the CAR process is suppressed in this regime. The CT conductance drops by increasing the length $L$, while the CAR conductance peaks at $L<\xi_{S}$. The charge conductance of the CT and CAR processes decay on the length scale of $k'^{-1}$ ($k')$ is the imaginary part of the longitudinal wave vector $k_j$ in the S region). In addition, there is an oscillatory behavior with $L/ \xi_{S}$ which pertains to the formation of the resonant transmission levels inside the S region (the oscillations can also be understood in the context of the Fabry-Perot oscillations~\cite{Born}). It is obvious from Figs.~\ref{Fig:4}(a)-\ref{Fig:4}(b) that the presence of the strong spin-orbit interaction and the $\beta$ term in the Hamiltonian of MoS$_2$ enhance the charge conductance of the CT and CAR processes, respectively, in the P and AP configurations, as compared with their value in the corresponding gapped graphene structure~\cite{note1}, and make them to decay slowly with the length $L/ \xi_{S}$, specially for the case of the $\beta=2.21$. Therefore, in contrast to the gapped graphene structure~\cite{note1}, the CT and CAR processes are present for long lengths of the S region in the MoS$_2$-based structure with $\beta=2.21$ and the CAR conductance peaks at a longer length of the S region. Figures~\ref{Fig:4}(c)-\ref{Fig:4}(d) show the behavior of the charge conductance of the CT (P configuration) and CAR (AP configuration) processes for the MoS$_2$-based structure with $\Delta= 1.9$, and $1.7$ eV. If we compare the results of $\Delta= 1.9$ eV with that of a smaller energy gap $\Delta= 1.7$ eV ($10$ percentage uncertainly can be achieved by exerting strain~\cite{Peelaers12}, for instance), we find that the charge conductance of the CT and CAR processes increase by decreasing the energy gap $\Delta$ and the corresponding process is present for longer lengths of the S region. Also, the period of the conductance oscillations for the case of the $\beta=2.21$ decreases by decreasing the energy gap $\Delta$ and the conductance of the CAR process peaks at a longer length of the S region.
\par
\begin{figure}[t]
\begin{center}
\includegraphics[width=3.5in]{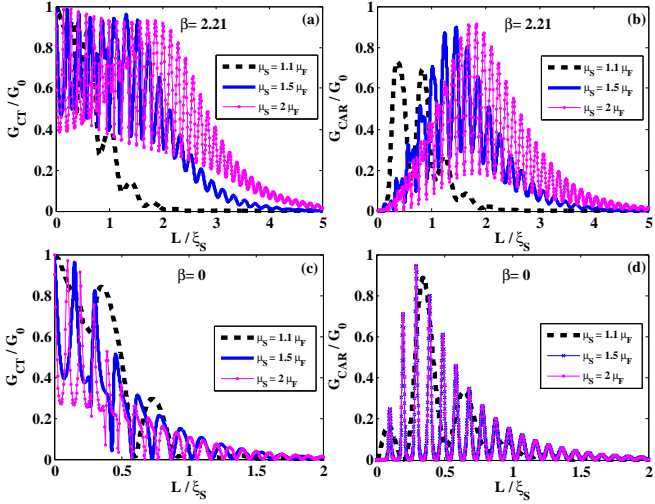}
\end{center}
\caption{\label{Fig:6} (Color online) Charge conductance of the CT and CAR processes as a function of $L/\xi_S$ for different values of the chemical potential $\mu_S$ in MoS$_2$-based structure with $\alpha=0$, $\beta=2.21$ (a)-(b) and $\beta=0$ (c)-(d), when $\Delta=1.9$ eV, $\mu_{F}=-\Delta/2+\lambda-h+eV+0.001$, $\Delta_S = 0.05$ eV, $h = 0.5 \lambda$ and $eV / \Delta_S = 0$.}
\end{figure}
Figure~\ref{Fig:5} shows the behavior of the charge conductance of the pure CT and pure CAR processes, respectively, in the P and AP configurations in terms of the length of the S region $L/\xi_S$ for two values of the bias voltage $eV/\Delta_S = 0$, and $1$, in the MoS$_2$-based structure with $\alpha=0$ and $\beta=0$ [Figs.~\ref{Fig:5}(a)-\ref{Fig:5}(b)] and $2.21$ [Figs.~\ref{Fig:5}(c)-\ref{Fig:5}(d)]. The charge conductance of the CT and CAR processes increase by increasing the bias voltage for $\beta=0$ [Figs. \ref{Fig:5}(a)-\ref{Fig:5}(b)] while for the case of the $\beta=2.21$, they can be increased or decreased depending on the length of the S region [Figs. \ref{Fig:5}(c)-\ref{Fig:5}(d)].
\begin{figure}[t]
\begin{center}
\includegraphics[width=3.45in]{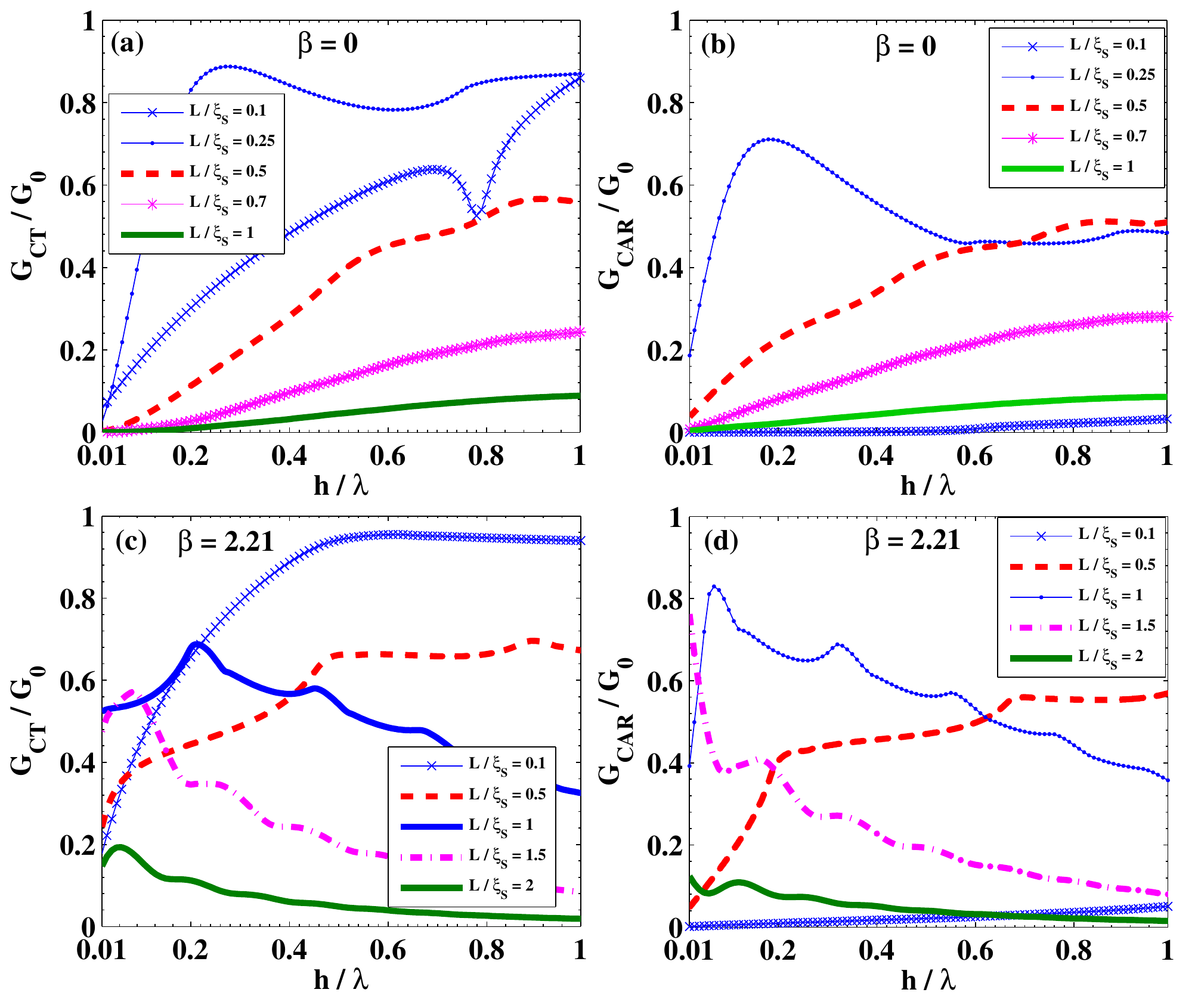}
\end{center}
\caption{\label{Fig:7} (Color online) Dependence of the charge conductance of the CT and CAR processes on the magnitude of the exchange field in the right F region $|\bm{h}_R|/\lambda=h/\lambda$ (in units of the spin-orbit coupling constant $\lambda$) for different lengths of the S region, when $\beta=0$ (a)-(b) and $2.21$ (c)-(d), $\alpha=0$, $eV/\Delta_S = 0$, $\mu_{F}=-\Delta/2+\lambda+eV-0.001$, $\mu_S = 1.1\ \mu_{F} = -1$ eV and $\Delta_S = 0.01$ eV.}
\end{figure}

Moreover, the behavior of the conductance of the pure CT and pure CAR processes versus $L/\xi_S$ is shown in Fig.~\ref{Fig:6} for different values of the chemical potential of the S region, $\mu_S$, in the MoS$_2$-based structure with $\alpha=0$ and $\beta=2.21$ [Figs. \ref{Fig:6}(a)-\ref{Fig:6}(b)] and $0$ [Figs. \ref{Fig:6}(c)-\ref{Fig:6}(d)], when $\Delta_S = 0.05$ eV. We consider the S region with the chemical potential $\mu_S>\mu_F$ to avoid the effect of the zigzag and armchair shapes at the interface. The conductance of the CT and CAR processes on the MoS$_2$-based structure with $\beta=2.21$ increase with $\mu_S$ and the corresponding process is present for long lengths of the S region, when $\mu_S$ is large. Moreover, the CAR conductance peaks at longer lengths of the S region and the period of oscillations in the conductance of the CT and CAR processes decrease by increasing the chemical potential $\mu_S$. Furthermore, if we compare the results of the case with $\beta=2.21$ and $\mu_S = 1.1\ \mu_{F}$ with those of the corresponding structure with $\Delta_S = 0.01$ eV [Figs. \ref{Fig:4}(a)-\ref{Fig:4}(b)], we find that the period of the oscillations increase by increasing $\Delta_S$. For the MoS$_2$-based structure with $\beta=0$ and large chemical potential $\mu_S$, the magnitude of the conductance of the CT and CAR processes strongly depend on $L/\xi_S$ value and importantly they can be zero for different lengths of the S region. In the case of the $\beta=2.21$, they have a finite value and can be controlled for different lengths of the S region, which makes this structure suitable for experimental measurements of the cross-conductance.

Figure~\ref{Fig:7} shows the behavior of the conductance of the pure CT and pure CAR processes in terms of the magnitude of the exchange field in the right F region $|\bm{h}_R|/\lambda=h/\lambda$ (in units of the spin-orbit coupling constant $\lambda$) for different lengths of the S region, when $\beta=0$ [Figs. \ref{Fig:7}(a)-\ref{Fig:7}(b)] and $2.21$ [Figs. \ref{Fig:7}(c)-\ref{Fig:7}(d)], $\alpha=0$, $eV/\Delta_S = 0$, $\mu_{F}=-\Delta/2+\lambda+eV-0.001$, $\mu_S = 1.1\ \mu_{F} = -1$ eV and $\Delta_S = 0.01$ eV. The conductance of the CT and CAR processes, increase or decrease with $h/\lambda$ value depending on the length of the S region. Comparing the results of the CT and CAR conductances, we find that the conductance of the CAR process can be smaller or larger than that of the CT process depending on the length of the S region and the exchange field $h/\lambda$.

Therefore, by studying the effect of different parameters on the charge conductance of the CT and CAR processes, respectively, in the P and AP alignments of magnetizations, we conclude that the suitable experimental set up to detect a fully valley- and spin-polarized nonlocal charge current in a superconducting spin valve may be a MoS$_2$-based structure (in the presence of the $\beta=2.21$) with smaller energy gap $\Delta=1.7$ eV, large chemical potentials $\mu_{F}$ [close to the energy of the valence band edge for $s=\tau=1$ spin-subband ($\mu_2$)] and $\mu_S$, in the presence of a large exchange field $h$ for $L<\xi_S$ and small $h$ for $L\geq\xi_S$. We note that in the proposed structure, we have considered F regions with the exchange field $h<\lambda$ while for the case of $h>\lambda$, the cross-conductance has a similar behavior for incoming electrons with $\mu_3 <\mu_{F}-\varepsilon\leq \mu_1$ ($\mu_3=-\Delta/2-\lambda+h$).
\begin{table*}[t!]
    \begin{center}
\begin{tabular}{l|c|c|c|r}
\hline\hline
\vline ~ & & & &  \vline\\
\vline ~ \textsl{Structure} & \textsl{Initial conditions} & \textsl{Charge current's polarization} & \textsl{Scattering processes} & \textsl{Advantages} \vline \\
\vline ~ & & & &  \vline\\
\hline
\vline ~ & & & &  \vline\\
\vline ~ N/S/N (Graphene)~\cite{Cayssol08}& $eV=\mu$ & ------ & Pure CAR & ------   \vline\\
\vline ~ &  & & & \vline\\
\vline ~ & & & & \vline\\
\hline
\vline ~ & & & & \vline\\
\vline ~ F/S/F (Graphene)~\cite{Linder09}& $\mu=h$ & "Spin-polarization"& Pure CT (P) & For all subgap  \vline\\
\vline ~ & & Spin-switch& & bias voltages\vline\\
\vline ~ & & between P and AP&Pure CAR (AP) & \vline\\
\vline ~ & & & & \vline\\
\hline
\vline ~ & & & &  \vline\\
\vline ~ MZR/S/ZR (Graphene)~\cite{JWang12}& Even zigzag  & "Spin-polarization"& Pure CAR & For wide range \vline\\
\vline ~ & chain number& Spin-switch & &of bias and  \vline\\
\vline ~  & for the ribbon & by changing  & & gate voltages \vline\\
\vline ~ & &the bias voltage &  & \vline\\
\vline ~ & & $eV_L>0\rightarrow eV_L<0$& & \vline\\
\hline
\vline ~ & & & & \vline\\
\vline ~ N/S/N (Silicene)~\cite{Linder14}& ------ & "Spin$\times$valley  & Pure CAR & For all subgap  \vline\\
\vline ~ & & polarization"  & & bias voltages\vline\\
\vline ~ & & ($s\tau=-1$)& & \vline\\
\hline
\vline ~ & & & &  \vline\\
\vline ~ F/S/F (MoS$_2$)& $\mu_2 <\mu-\varepsilon\leq \mu_1$ & "Spin-polarization" & Pure CT (P)& 1) Without fixing \vline\\
\vline ~ & ($h<\lambda$)& and & & of any parameter,\vline\\
\vline ~ & &"Valley-polarization" &Pure CAR (AP) &for a wide range \vline\\
\vline ~ & $\mu_3 <\mu-\varepsilon\leq \mu_1$& & & of chemical potential\vline\\
\vline ~ & ($h>\lambda$)& Spin-switch& &  and subgap bias voltage\vline\\
\vline ~ & & and& &   \vline\\
\vline ~ &  & valley-switch& &2) Applicable in  \vline\\
\vline ~ & & between P and AP & & valleytronics\vline\\
\vline ~ & & & &  \vline\\
\vline ~ & & & &3) Enhanced and\vline\\
\vline ~ & & & & long-range G in \vline\\
\vline ~ & & & &presence of $\beta$ and $\lambda$  \vline\\
\hline\hline
\end{tabular}
\caption{ Properties of the atomically thin two-dimensional crystals-based proposed structures for detecting pure CAR and pure CT processes. The results of the graphene-based N/S/N, F/S/F and MZR/S/ZR structures are, respectively, reported from Refs.~[\onlinecite{Cayssol08}], [\onlinecite{Linder09}] and [\onlinecite{JWang12}]. The results of the silicene-based structure is reported from Ref.~[\onlinecite{Linder14}]. P(AP) stands for the parallel (antiparallel) alignment of magnetizations in spin valve (F/S/F) structures. $V_L$ is the bias voltage, applied on the left ZR. $\mu_1=-\Delta/2+\lambda+h$, $\mu_2=-\Delta/2+\lambda-h$ and $\mu_3=-\Delta/2-\lambda+h$ are the energies of the valence band edges for different spin-subbands of two valleys.}
\label{table1}
\end{center}
\end{table*}

\subsection{\label{subsecB}Thermal Conductance}
\begin{figure}[t]
\begin{center}
\includegraphics[width=3.5in]{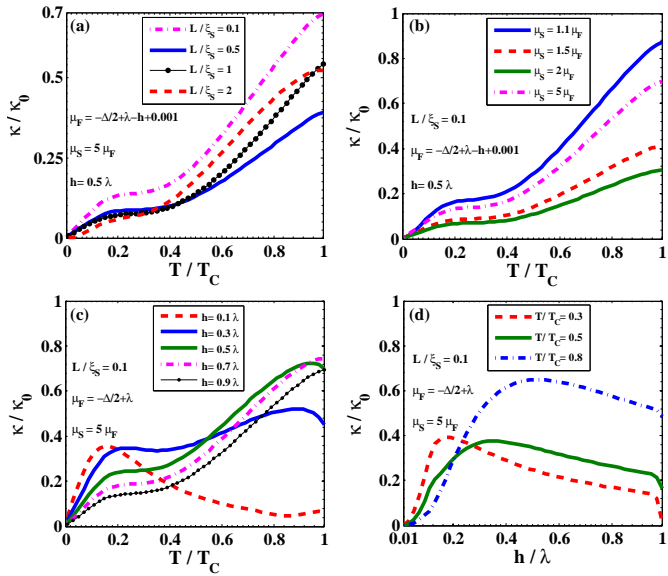}
\end{center}
\caption{\label{Fig:8} (Color online) Thermal conductance of the MoS$_2$-based F/S/F structure with parallel alignment of magnetizations as a function of $T/T_C$ for different lengths and chemical potentials of the S region, when $\mu_F=-\Delta/2+\lambda-h+0.001$, and $h=0.5 \lambda$ (a)-(b) and for different values of the exchange field in F region, when $L/\xi_S = 0.1$, $\mu_F=-\Delta/2+\lambda$ and $\mu_S = 5\ \mu_F$ (c). (d) The thermal conductance as a function of $h/\lambda$ for different values of $T/T_C$, when $L/\xi_S = 0.1$, $\mu_F=-\Delta/2+\lambda$ and $\mu_S = 5\ \mu_F$.}
\end{figure}
Now, we proceed to investigate the thermal transport characteristics of the MoS$_2$-based F/S/F structure with parallel alignment of magnetizations. We replace the zero-temperature superconducting order parameter $\Delta_S$ in Eq. (\ref{DBdG}) with the temperature-dependent one, $\Delta'_S(T)=1.76 k_B T_C\tanh{(1.74\sqrt{{T_C}/{T}-1})}$. As before, we consider the chemical potential of the F region, $\mu_F$, in the range $\mu_2 <\mu_{F}< \mu_1$ by assuming that the low-energy excitations are excited in the system. We scale the temperature, T, in units of the critical temperature of the superconducting order parameter, $T_C$, and we set $k_B=1$, $\alpha=0$ and $\beta=2.21$ throughout our computations.

Figures \ref{Fig:8}(a)-\ref{Fig:8}(b) show the behavior of the normalized thermal conductance $\kappa/\kappa_0$ ($\kappa_0=2A'\sum_{s,\tau=\pm1}\int_0^{\infty} d\varepsilon\ |\bm{k}_{e}^{s,\tau}(\varepsilon)|$) in terms of the temperature $T/T_C$ for various lengths and chemical potentials of the S region, when $h=0.5 \lambda$ and $\mu_F = -\Delta/2+\lambda-h+0.001$. We find that for small lengths of the highly doped S region, the thermal conductance has a linear behavior with respect to $T/T_C$ at low temperatures \textit{i.e.} $\kappa\propto T$, which is similar to the thermal conductance of metals~\cite{Bardas95,Devyatov}, and exponential dependence on $T/T_C$ at high temperatures \textit{i.e.} $\kappa\propto [a\ exp(b T)+c\ exp(d T)]$. As clear from Fig. \ref{Fig:8}(b), decreasing the value of the chemical potential of the S region, $\mu_S$, leads to the suppression of the linear and the exponential dependence of the thermal conductance on the temperature. Figure \ref{Fig:8}(c) displays the temperature dependence of the thermal conductance for different values of the exchange field, $h$, in the F/S/F structure with a small length of the highly doped S region ($L/\xi_S=0.1$, $\mu_S= 5\ \mu_F$), when $\mu_F= -\Delta/2+\lambda$. The thermal conductance has different behavior with respect to the temperature, depending on the value of the exchange field. While it has an increasing behavior with $T$ for large values of $h$, it can be increased or decreased with $T$ for small values of $h$. We further demonstrate the exchange field dependence of the thermal conductance for different values of $T/T_C$ in Fig. \ref{Fig:8}(d). The thermal conductance increases with the exchange field and attains a maximum value at $h<\lambda<|\mu_F|$, which tends toward to higher exchange fields by increasing the temperature. This behavior of the thermal conductance is in contrast to that of a graphene-based F/S/F structure~\cite{Salehi10}, where $\kappa$ decreases with the exchange field for $h<E_F$ and attains a minimum value near $h\simeq E_F$. Moreover, we find that the thermal conductance shows a damped oscillatory behavior with respect to the length of the S region, $L/\xi_S$, which is the consequence of the resonant transmission levels inside the S region.

\section{\label{sec:level3}Conclusion}
We have investigated the charge and thermal transport characteristics of a $p$-doped molybdenum disulfide ferromagnetic/superconducting/ferromagnetic (F/S/F) junction which constitutes a superconducting spin valve structure. We have found valley- and spin-switching effects, without fixing of any parameter and for a determined range of the chemical potential of the F region in which the subgap transport of electrons can be switched from a purely elastic electron cotunneling (CT) process to a pure crossed Andreev reflection (CAR) process by changing the alignment of the magnetizations of F regions from parallel to antiparallel. This makes the nonlocal charge current to be fully valley- and spin-polarized inside the right F region, and the type of the polarizations can be changed by reversing the magnetization direction in the right F region. We have demonstrated that the presence of the strong spin-orbit interaction and the topological term ($\beta$) in the Hamiltonian of MoS$_2$ enhance the charge conductance of the CT and CAR processes, respectively, in the parallel and antiparallel configurations and make them to decay slowly with the length of the S region.

Since several works on the CT and CAR processes in two-dimensional systems are available, a proper comparison with those results seems to be in order ( see Table 1). Cayssol~\cite{Cayssol08} focused on graphene-based N/S/N structure at the bias voltage $eV=\mu$ and predicted a pure CAR process without any valley- or spin-polarization. Linder~{\it et al.}~\cite{Linder09}, on the other hand, proposed a graphene-based superconducting spin valve (F/S/F) structure which creates a spin-polarized nonlocal current via the pure CAR (CT) process in the antiparallel (parallel) configuration of the magnetizations of F regions at a fixed chemical potential $\mu=h$. In these structures, a precise bias or gate voltage is needed to modulate the local Fermi energy at the Dirac point and suppress both CT (CAR) and AR processes to have a pure CAR (CT) process. However, any energy fluctuation may lead to deviation from the Dirac point and a sizable CT (CAR) transmission. A pure CAR process has also been shown in a magnetized zigzag graphene nanoribbon (MZR)/S/ZR junction with an even zigzag chain number for the ribbon~\cite{JWang12}. Furthermore, Linder \textit{et al.}~\cite{Linder14} recently demonstrated a fully spin$\times$valley polarized pure CAR process in silicene-based N/S/N structure which means that the nonlocal current is fully spin-polarized in each valley. It would be worthwhile mentioning that in the above structures, fixing of a unique parameter is needed to have a pure CAR (CT) process. Also, the nonlocal current through these structures may be spin- (spin$\times$valley-) polarized. Interestingly, we have found that in the MoS$_2$-based structure, there is no fixing of a unique parameter and the nonlocal charge current has full valley-polarization in addition to spin-polarization, which makes this structure applicable in valleytronics. We have further argued that the suitable experimental set up may be a MoS$_2$-based structure with smaller energy gap $\Delta$ and large chemical potentials in the S and F regions (the chemical potential in the F region should be close to the energy of the valence band edges for $s=\tau= 1$ spin-subbands).

Furthermore, we have presented the investigation of the thermal transport in the structure. We have investigated the effect of the variation of various parameters on the thermal conductance of the structure and found that for small lengths of the highly doped S region, the thermal conductance displays linear dependence on temperature at low temperatures and exponential increase versus temperature at high temperatures. The behavior of the thermal conductance also depends on the strength of the exchange field in the F region such that the thermal conductance can be increased or decreased with temperature, depending on the value of the exchange field. We have further demonstrated that the thermal conductance versus the exchange field has a maximum value at $h<\lambda<|\mu_F|$, which goes towards larger exchange fields by increasing the temperature.
\begin{acknowledgements}
We would like to thank G. E. W. Bauer for useful discussions. R. A. would like to thank the Victoria University of Wellington for its hospitality during the period when the last part of this work was carried out.
\end{acknowledgements}
\appendix*
\section {\label{sec:appendix} Derivation of the Dirac-Bogoliubov-de Gennes (DBdG) equation for monolayer MoS$_2$}
By introducing an effective mean-field Hamiltonian and using the Bogoliubov transformations, the Bogolibov-de Gennes (BdG) equation can be obtained as
\begin{eqnarray}
&&\hspace{1.5cm}H_{BdG}\left(
\begin{array}{c}
u\\
v
\end{array}
\right)
=\varepsilon\left(
\begin{array}{c}
u\\
v
\end{array}
\right),\nonumber\\\nonumber\\
\label{BdG}
&&\left(
\begin{array}{cc}
\mathcal{H}-\mu & \Delta_S \\
\Delta_{S}^{\ast}& \mu-\mathcal{T}\mathcal{H}\mathcal{T}^{-1}
\\
\end{array}
\right)
\left(
\begin{array}{c}
u\\
v
\end{array}
\right)
=\varepsilon\left(
\begin{array}{c}
u\\
v
\end{array}
\right),
\end{eqnarray}
where $\mathcal{H}$ is the effective single-particle Hamiltonian and $\mathcal{T}$ is the time-reversal operator.

The effective single-particle Hamiltonian of monolayer MoS$_2$, where we ignore the intravalley interaction, has the form
\begin{equation}
\label{Htau}
\mathcal{H}=\left(
    \begin{array}{cc}
      \mathcal{H}_{\tau} & 0 \\
      0 & \mathcal{H}_{\bar{\tau}} \\
    \end{array}
  \right),
\end{equation}
with
\begin{eqnarray}
\label{H}
H_{\tau}&=&v_{\rm F}(\bm{\sigma}_{\tau}.\bm{p})+\frac{\Delta}{2}\sigma_{z}+\lambda s \tau\ (\frac{1-\sigma_z}{2})\nonumber\\
&+&\frac{\bm{p}^2}{4m_0}(\alpha+\beta\sigma_z)+U(\bm{r}),
\end{eqnarray}
for valley $\tau=\pm 1$ ($\bar{\tau}=-\tau$). This effective Hamiltonian $\mathcal{H}$ is time-reversal invariant $\mathcal{T}\mathcal{H}(\bm{p})\mathcal{T}^{-1}=\mathcal{H}(-\bm{p})$ with $\mathcal{T}=i\tau_x s_y \mathcal{K}$ ($\mathcal{K}$ is the operator of complex conjugation).

The BdG Hamiltonian should have particle-hole symmetry, which requires that for each eigenstate $\Psi(\bm{p})$ of the Hamiltonian with energy $\varepsilon(\bm{p})$, there is another eigenstate $\mathcal{T}C\Psi(\bm{p})$ with energy $-\varepsilon(-\bm{p})$ ($\mathcal{T}C$ is the operator of particle-hole symmetry). We suppose that ${(u(\bm{p}),v(\bm{p}))}^T$ is the eigenstate of Bogoliubov Hamiltonian $H_{BdG}(\bm{p})$ with energy $\varepsilon(\bm{p})$. Using Eq. (\ref{BdG}) and $\mathcal{T}\mathcal{H}(\bm{p})\mathcal{T}^{-1}=\mathcal{H}(-\bm{p})$, we obtain
\begin{widetext}
\begin{equation}
\label{bdg2}
\hspace{-3mm}
\left(
\begin{array}{cc}
\mathcal{H}(\bm{p})-\mu & \Delta_S \\
\Delta_S^{\ast}& \mu-\mathcal{H}(-\bm{p})
\\
\end{array}
\right)
\left(
\begin{array}{c}
\mathcal{T}v(\bm{p})\\
-\mathcal{T}u(\bm{p})
\end{array}
\right)
=-\varepsilon(-\bm{p})\left(
\begin{array}{c}
\mathcal{T}v(\bm{p})\\
-\mathcal{T}u(\bm{p})
\end{array}
\right).
\end{equation}
\end{widetext}
This tells us that
\begin{equation}
\left(
\begin{array}{c}
\mathcal{T}v(\bm{p})\\
-\mathcal{T}u(\bm{p})
\end{array}
\right)=\mathcal{T}i\gamma_y\left(
\begin{array}{c}
u(\bm{p})\\
v(\bm{p})
\end{array}\right)
\end{equation}
is an eigenstate of BdG Hamiltonian $H_{BdG}(\bm{p})$, with energy $-\varepsilon(-\bm{p})$ and therefore the particle-hole operator is $\mathcal{T}C$, with $C=i\gamma_y$ (the Pauli matrix $\gamma_y$ acts on the electron-hole space). Using Eqs. (\ref{BdG}) and (\ref{bdg2}), we find that $H_{BdG}(\bm{p})$ anticommutes with the operator of particle-hole symmetry ($\mathcal{T}C$) \begin{equation}
H_{BdG}(\bm{p})\ \mathcal{T}C+\mathcal{T}C H_{BdG}(\bm{p})=0,
\end{equation}
in which $C H_{BdG}(\bm{p})C^{-1}=-H_{BdG}(-\bm{p})$.

Substituting the time-reversal operator $\mathcal{T}$ into Eq. (\ref{BdG}), results in two decoupled sets of four-dimensional Dirac-Bogoliubov-de Gennes (DBdG) equations, which each of the form is given by
\begin{equation}
\label{DBdG_ap}
\left(
\begin{array}{cc}
\mathcal{H}_{\tau}-\mu & \Delta_S \\
\Delta_{S}^{\ast}& \mu-\mathcal{H}_{\tau}
\\
\end{array}
\right)
\left(
\begin{array}{c}
u_{\tau}\\
v_{\bar{\tau}}
\end{array}
\right)
=\varepsilon\left(
\begin{array}{c}
u_{\tau}\\
v_{\bar{\tau}}
\end{array}
\right).
\end{equation}

In the presence of an exchange field $\bm{h}=h\hat{z}$, the Hamiltonian of monolayer MoS$_2$ takes the form
\begin{equation}
\acute{H}(\bm{p})=\mathcal{H}(\bm{p}) \hat{I}_{2\times2}-h\tau_0 s_{z},
\end{equation}
which acts on the wave function $\Psi(\bm{p})={(u(\bm{p}),v(\bm{p}))}^T$ with $u(\bm{p})=(u_{s,\tau},u_{\bar{s},\tau}, u_{s,\bar{\tau}},u_{\bar{s},\bar{\tau}})^T$ and  $v(\bm{p})=(v_{s,\tau},v_{\bar{s},\tau}, v_{s,\bar{\tau}},v_{\bar{s},\bar{\tau}})^T$. It can be shown that
\begin{equation}
\label{HF}
\mathcal{T}{\acute{H}}(\bm{p})\mathcal{T}^{-1}=\mathcal{H}(-\bm{p}) \hat{I}_{2\times2}+h\tau_0 s_{z}.
\end{equation}

Replacing Eqs. (\ref{HF}) and (\ref{Htau}) in BdG equation [Eq. (\ref{BdG})], we find that the DBdG equation in presence of an exchange interaction can be decoupled to four sets of equations, which each of the form is given by
\begin{equation}
\left(
\begin{array}{cc}
H_{\tau}-sh & \Delta_S \\
\Delta_{S}^{\ast}& -(H_{\tau}-{\bar{s}}h)
\\
\end{array}
\right)
\left(
\begin{array}{c}
u_{s,\tau}\\
v_{\bar{s},\bar{\tau}}
\end{array}
\right)
=\varepsilon\left(
\begin{array}{c}
u_{s,\tau}\\
v_{\bar{s},\bar{\tau}}
\end{array}
\right).
\end{equation}
This is the Eq. (\ref{DBdG}) in Sec. \ref{sec:level1}.


\begin{thebibliography}{99}
\bibitem{note1}
We compare the results of the MoS$_2$-based structure ($\lambda=0.08$ eV, $\beta=2.21$, $\Delta=1.9$ eV) with those of the introduced gapped graphene structure with very large energy gap ($\lambda=0$, $\beta=0$, $\Delta=1.9$ eV).
\end{thebibliography}
\end{document}